\documentclass[eqsecnum,pra,aps,preprint,superscriptaddress]{revtex4}

\usepackage{amsmath}
\usepackage{epsfig}
\begin{document}

\title{On the torque on birefringent plates induced by quantum fluctuations}
\author{Davide Iannuzzi}
\thanks{The two authors equally contributed to this work}
\affiliation{Harvard University, Division of Engineering and Applied Sciences, Cambridge, MA 02138}
\author{Jeremy N. Munday}
\thanks{The two authors equally contributed to this work}
\affiliation{Harvard University, Department of Physics, Cambridge, MA 02138}
\author{Yuri Barash}
\affiliation{Institute of Solid State Physics, Russian Academy of Sciences, Institutskaya ul. 2, Chernogolovka, Moscow, reg. 142432 , Russia}
\author{Federico Capasso}
\email{capasso@deas.harvard.edu}
\affiliation{Harvard University, Division of Engineering and Applied Sciences, Cambridge, MA 02138}

\begin{abstract}
We present detailed numerical calculations of the mechanical torque induced by quantum fluctuations on two parallel birefringent plates with in plane optical anisotropy, separated by either vacuum or a liquid (ethanol). The torque is found to vary as $\sin(2\theta)$, where $\theta$ represents the angle between the two optical axes, and its magnitude rapidly increases with decreasing plate separation $d$. For a 40 $\mu$m diameter disk, made out of either quartz or calcite, kept parallel to a Barium Titanate plate at $d\simeq 100$ nm, the maximum torque (at $\theta={\pi\over 4}$) is of the order of $\simeq 10^{-19}$ N$\cdot$m. We propose an experiment to observe this torque when the Barium Titanate plate is immersed in ethanol and the other birefringent disk is placed on top of it. In this case the retarded van der Waals (or Casimir-Lifshitz) force between the two birefringent slabs is repulsive. The disk would float parallel to the plate at a distance where its net weight is counterbalanced by the retarded van der Waals repulsion, free to rotate in response to very small driving torques.

\end{abstract}

\maketitle

\section{INTRODUCTION}

According to quantum electrodynamics, quantum fluctuations of electric and magnetic fields give rise to a zero-point energy that never vanishes, even in the absence of electromagnetic sources\cite{milonni}. In 1948, H. B. G. Casimir predicted that, as a consequence, two electrically neutral metallic parallel plates in vacuum, assumed to be perfect conductors, should attract each other with a force inversely proportional to the fourth power of separation\cite{casimir}. The plates act as a cavity where only the electromagnetic modes that have nodes on both the walls can exist. The zero-point energy (per unit area) when the plates are kept at close distance is smaller than when the plates are at infinite separation. The plates thus attract each other to reduce the energy associated with the fluctuations of the electromagnetic field.

E. M. Lifshitz, I. E. Dzyaloshinskii, and L. P. Pitaevskii generalized Casimir's theory to isotropic dielectrics \cite{lifshitz1,lifshitz2,lifshitz3}. In their theory the force between two uncharged parallel plates with arbitrary dielectric functions can be derived according to an analytical formula that relates the Helmholtz free energy associated with the fluctuations of the electromagnetic field to the dielectric functions of the interacting materials and of the medium in which they are immersed\cite{noteonlif2}. At very short distances (typically smaller than a few nanometers), Lifshitz's theory provides a complete description of the non-retarded van der Waals force. At larger separations, retardation effects give rise to a long-range interaction that in the case of two ideal metals in vacuum reduces to Casimir's result. Lifshitz's equation also shows that two plates made out of the same material always attract, regardless of the choice of the intervening medium. For slabs of different materials, on the contrary, the sign of the force depends on the dielectric properties of the medium in which they are immersed\cite{ninham,isra}. While the force is always attractive in vacuum, there are situations for which a properly chosen liquid will cause the two plates to repel each other\cite{ninham,isra}. 

As mentioned above, one of the limitations of Lifshitz's theory\cite{lifshitz1} is the assumption that the dielectric properties of the interacting materials are isotropic. In 1972 V. A. Parsegian and G. H. Weiss derived an equation for the non-retarded van der Waals interaction energy between two dielectrically anisotropic plates immersed in a third anisotropic material\cite{parse}. One of the authors of the present paper (Y. B.) analyzed a similar problem and found an equation for the Helmholtz free energy (per unit area) of the electromagnetic field in which retardation effects are included\cite{barash}. In the non-retarded limit, the two results are in agreement.

Both articles also show that a torque develops between two parallel birefringent slabs (with in plane optical anisotropy, as shown in figure \ref{Barash_config}) placed in a isotropic medium, causing them to spontaneously rotate towards the configuration in which their principal axes are aligned. This effect can be qualitatively understood by noting that the relative rotation of the two plates will result in a modification of the zero-point energy, because the reflection, transmission, and absorption coefficients of these materials depend on the angle between the wave vector of the virtual photons, responsible for the zero-point energy, and the optical axis. The anisotropy of the zero-point energy between the plates then generates the torque that makes them rotate toward configurations of smaller energy.

The Casimir-Lifshitz force between isotropic dielectrics is receiving considerable attention in the modern literature. The theory has been verified in several high precision experiments, and although the investigation has been mainly focused on the interaction between metallic surfaces in vacuum, there are no doubts about its general validity\cite{hochan1,over,lamo,moh1,moh2,moh3,ederth,bressi,decca} (for a review of previous measurements, see \cite{isra,repmoh}; for a critical discussion on the precision of the most recent experiments, see \cite{iannuzziconf,svetovoyconf,mohpra,milton}). Less precise measurements in liquids have been reported\cite{isra,suresh,prieve}, and experimental evidence for repulsive van der Waals forces between dielectric surfaces in different fluids has also been reported\cite{repulsive1,repulsive2,repulsive3}. Finally, it has been pointed out that the Casimir-Lifshitz force might be a potentially relevant issue for the development of micro- and nanoelectromechanical systems\cite{hochan1,serry,hochan2}. 

On the other hand, essentially no attention has been devoted to the torque between anisotropic materials predicted by Parsegian, Weiss and Barash, with the exception of a theoretical derivation of a more simplified equation of the torque between two plates in a one dimension calculation\cite{vanenk} and between engineered anisotropic surfaces (two ellipsoids with anisotropic dielectric function\cite{bookmoste} and two dielectric slabs with different directions of conductivity\cite{kenneth}). No experimental attempts to demonstrate the effect have ever been reported, and so far no numerical calculations to estimate its magnitude have been presented\cite{prieve2}. 

In this paper we calculate the magnitude of the torque induced by quantum fluctuations for specific materials and discuss possible experimental validations of the effect. We consider a small birefringent disk (diameter 40 $\mu$m, thickness 20 $\mu$m) made out of either quartz or calcite placed parallel to a birefringent Barium Titanate (BaTiO$_3$) plate in vacuum. Using the dielectric properties for the materials reported in the literature, we show that the magnitude of the torque is within the sensitivity of available instrumentation, provided that the plate and the disk are kept at sub-micron distances. Unfortunately, at such short separations the tendency of the two surfaces to stick together represents a major technical difficulty. Therefore, the measurement of the rotation of the disk may seem to be an extremely challenging problem, which can be only addressed with the design of sophisticated mechanical systems. In this paper we propose a much simpler experimental approach, where the BaTiO$_3$ plate is immersed in liquid ethanol and the quartz or calcite disk is placed on top of it. In this case the retarded van der Waals force between the two birefringent slabs is repulsive. The disk is thus expected to float on top of the plate at a distance of approximately 100 nm, where its weight is counterbalanced by the van der Waals repulsion. Because there is no contact between the two birefringent surfaces, the disk would be free to rotate in a sort of \emph{frictionless bearing}, sensitive to even very small driving torques.

\section{THEORY}

Let's consider two plates made of uniaxial birefringent materials kept parallel at a distance $d$ and immersed inside a medium with dielectric function $\epsilon_3$. For sake of simplicity, let's assume that the two plates are oriented as in figure \ref{Barash_config}. The $z$-axis of our reference system is chosen to be orthogonal to the plates. The optical axis of one of the two crystals (i.e. the threefold, fourfold, or sixfold axis of symmetry for rhombohedral, tetragonal, or hexagonal crystals, respectively\cite{lifshitz4}) is aligned with the $x$-axis. The optical axis of the second crystal is also in the $x-y$ plane but rotated by an angle $\theta$ with respect to the other. The dielectric tensors of the two plates are then described by the following matrices\cite{lifshitz4}:

\begin{equation}
\begin{pmatrix}
\epsilon_{1\|} & 0 & 0 \\
0 & \epsilon_{1\bot} & 0 \\
0 & 0 & \epsilon_{1\bot} 
\end{pmatrix}
\qquad , \qquad
\begin{pmatrix}
\epsilon_{2\|}\cos^2\theta+\epsilon_{2\bot}\sin^2\theta & & &
(\epsilon_{2\bot}-\epsilon_{2\|})\sin\theta\cos\theta & & & 0 \\
(\epsilon_{2\bot}-\epsilon_{2\|})\sin\theta\cos\theta & & &
\epsilon_{2\|}\sin^2\theta+\epsilon_{2\bot}\cos^2\theta & & & 0 \\
0 & & & 0 & & & \epsilon_{2\bot}
\label{ematr}
\end{pmatrix}
\end{equation}

\noindent where the subscripts $\|$ and $\bot$ indicate the value of the dielectric tensor along the optical axis and along the plane orthogonal to the optical axis, respectively. It is important to stress that in equation \ref{ematr} $\epsilon_{1\|}$, $\epsilon_{1\bot}$, $\epsilon_{2\|}$, and $\epsilon_{2\bot}$ are functions of the angular frequency of the electromagnetic wave $\omega$.

The Helmholtz free energy (per unit area) of the system is given by\cite{barash}:

\begin{equation}
\Omega(\theta,d)={k_BT\over 4\pi^2}\sideset{}{'}\sum_{n=0}^\infty\int_0^\infty r dr \int_0^{2\pi} d{\varphi} \ln D_n(\varphi,r)
\label{energy}
\end{equation}

\noindent where $k_B$ is the Boltzmann constant, $T$ is the temperature of the system, and

\begin{equation}
\begin{split}
 D_n(\varphi,r)=
&
\cfrac{1}{\gamma} \cdot \biggl\{
 A-
 {(\widetilde{\rho}_2-\rho_2)\epsilon_{2\bot}^{(n)} \over \rho_2^2 -r^2\sin^2(\phi+\theta)} \medspace \\
&
\cdot \bigl\{ Br^2\sin^2(\phi+\theta)-E[2r^2\sin\varphi\cos\theta\sin(\phi+\theta)
+\rho_3^2\sin^2\theta]+C \bigr\} \biggr\} \\
\end{split}
\label{D}
\end{equation}

\begin{equation}
\begin{split}
\gamma =
& (\rho_1+\rho_3)(\rho_2+\rho_3) \cdot \biggl[ (\epsilon_3^{(n)}\rho_1+\epsilon_{1\bot}^{(n)}\rho_3)-{\epsilon_{1\bot}^{(n)}
(\widetilde{\rho}_1
-\rho_1)(r^2\sin^2\varphi - \rho_1\rho_3) \over \rho_1^2 - r^2\sin^2\varphi} \biggr] \\
&
\cdot \biggl[ (\epsilon_3^{(n)}\rho_2+\epsilon_{2\bot}^{(n)}\rho_3)-
{\epsilon_{2\bot}^{(n)}(\widetilde{\rho}_2-\rho_2)[(r\cos\varphi\sin\theta+r\sin\varphi\cos\theta)^2
-\rho_2\rho_3] \over \rho_2^2 (r\cos\varphi\sin\theta+r\sin\varphi\cos\theta)^2} \biggr] \\
\end{split}
\label{eqgamma}
\end{equation}

\begin{equation}
\begin{split}
A=
&
[(\rho_1+\rho_3)(\rho_2+\rho_3)-(\rho_1-\rho_3)(\rho_2-\rho_3)
\exp(-2\rho_3d)] \\
&
\cdot [(\epsilon_3^{(n)}\rho_1+\epsilon_{1\bot}^{(n)}\rho_3)(\epsilon_3^{(n)}\rho_2+
\epsilon_{2\bot}^{(n)}\rho_3)-(\epsilon_3^{(n)}\rho_1-\epsilon_{1\bot}^{(n)}\rho_3)(\epsilon_3^{(n)}\rho_2-
\epsilon_{2\bot}^{(n)}\rho_3)\exp(-2\rho_3d)] \\
&
-{(\widetilde{\rho_1}-\rho_1)\epsilon_{1\bot}^{(n)}\over 
\rho_1^2-r^2\sin^2\varphi}\{(r^2\sin^2\varphi-\rho_1\rho_3)
(\epsilon_3^{(n)}\rho_2+\epsilon_{2\bot}^{(n)}\rho_3)(\rho_2+\rho_3)(\rho_1+\rho_3) \\
&
+2(\epsilon_{2\bot}^{(n)}-\epsilon_3^{(n)})[r^2\sin^2\varphi (r^2\rho_1-\rho_2\rho_3^2)+\rho_1\rho_3^2(r^2-2r^2\sin^2\varphi+\rho_1\rho_2)]  \\
&
\cdot \exp(-2\rho_3d)+(r^2\sin^2\varphi+\rho_1\rho_3)
(\epsilon_3^{(n)}\rho_2-\epsilon_{2\bot}^{(n)}\rho_3)(\rho_1-\rho_3)(\rho_2-\rho_3)\exp(-4\rho_3d)\}
\\
\end{split}
\label{eqA}
\end{equation}

\begin{equation}
\begin{split}
B=
&
[(\epsilon_3^{(n)}\rho_1+\epsilon_{1\bot}^{(n)}\rho_3)(\rho_1+\rho_3)(\rho_2+\rho_3)
+2(\epsilon_{1\bot}^{(n)}-\epsilon_3^{(n)})(r^2\rho_2-\rho_1\rho_3^2-2\rho_2\rho_3^2)
\exp(-2\rho_3d) \\
&
+ (\epsilon_3^{(n)}\rho_1-\epsilon_{1\bot}^{(n)}\rho_3)(\rho_1-\rho_3)(\rho_2-\rho_3)\exp(-4\rho_3d)] \\
&
+{(\widetilde{\rho}_1-\rho_1)\epsilon_{1\bot}^{(n)} \over \rho_1^2-r^2\sin^2\varphi}
\{-(r^2\sin^2\varphi-\rho_1\rho_3)(\rho_1+\rho_3)(\rho_2+\rho_3)
\\
&
+2[r^2\sin^2\varphi(\rho_1\rho_2+\rho_3^2)-\rho_1^2\rho_3^2+\rho_1\rho_2\rho_3^2]
\exp(-2\rho_3d)\\
&
-(r^2\sin^2\varphi+\rho_1\rho_3)(\rho_1-\rho_3)(\rho_2-\rho_3)\exp(-4\rho_3d)\}
\\
\end{split}
\label{eqB}
\end{equation}

\begin{equation}
\begin{split}
C=
&
\rho_2\rho_3[-(\epsilon_3^{(n)}\rho_1+\epsilon_{1\bot}^{(n)}\rho_3)(\rho_1+\rho_3)(\rho_2+\rho_3)
+2\rho_3(\epsilon_{1\bot}^{(n)}-\epsilon_3^{(n)}) \\
&
\cdot (r^2+\rho_1\rho_2)\exp(-2\rho_3d)+(\epsilon_3^{(n)}\rho_1-
\epsilon_{1\bot}^{(n)}\rho_3)(\rho_1-\rho_3)(\rho_2-\rho_3)\exp(-4\rho_3d)] \\
&
+{(\widetilde{\rho_1}-\rho_1)\epsilon_{1\bot}^{(n)}\over 
\rho_1^2-r^2\sin^2\varphi}\rho_2\rho_3\{(r^2\sin^2\varphi-\rho_1\rho_3)
(\rho_1+\rho_3)(\rho_2+\rho_3) \\
&
+2\rho_3[\rho_1^2\rho_2+\rho_1\rho_3^2+r^2\sin^2\varphi(\rho_1-\rho_2)]
\exp(-2\rho_3d) \\
&
-(r^2\sin^2\varphi+\rho_1\rho_3)(\rho_1-\rho_3)(\rho_2-\rho_3)\exp(-4\rho_3d)\}
\label{eqC}
\end{split}
\end{equation}

\begin{equation}
E=4\rho_1\rho_2\rho_3^2
\cdot{(\widetilde{\rho_1}-\rho_1)\epsilon_{1\bot}^{(n)}\over 
\rho_1^2-r^2\sin^2\varphi}\exp(-2\rho_3d)
\label{eqE}
\end{equation}

\begin{equation}
\rho_1^2=r^2+{\xi_n^2 \over c^2}\epsilon_{1\bot}^{(n)}
\quad , \quad
\rho_2^2=r^2+{\xi_n^2 \over c^2}\epsilon_{2\bot}^{(n)}
\quad , \quad
\rho_3^2=r^2+{\xi_n^2 \over c^2}\epsilon_3^{(n)}
\label{eqrho3}
\end{equation}

\begin{equation}
\widetilde{\rho}^2_1=r^2+\biggl({\epsilon_{1\|}^{(n)}\over \epsilon_{1\bot}^{(n)}}-1\biggr)r^2\cos^2\varphi+{\xi_n^2 \over c^2}
\epsilon_{1\|}^{(n)}
\end{equation}
\begin{equation}
\widetilde{\rho}^2_2=r^2+\biggl({\epsilon_{2\|}^{(n)}\over \epsilon_{2\bot}^{(n)}}-1\biggr)(r\cos\varphi\cos\theta-r\sin\varphi\sin\theta)^2+{\xi_n^2 \over c^2}\epsilon_{2\|}^{(n)}
\label{eqrho2t}
\end{equation}

\noindent where $c$ is the speed of light in vacuum. The prime in the summation in equation \ref{energy} indicates that the first term ($n=0$) must be multiplied by a factor ${1\over 2}$. Furthermore $\epsilon_{1\|}^{(n)}$, $\epsilon_{1\bot}^{(n)}$, $\epsilon_{2\|}^{(n)}$, $\epsilon_{2\|}^{(n)}$, and $\epsilon_3^{(n)}$ represent the values of the dielectric functions of the interacting materials calculated at imaginary angular frequencies\cite{lifshitz4} $i\xi_n=i\cdot {2 \pi k_B T \over \hbar} \cdot n$, where $\hbar$ is the Planck constant divided by $2\pi$.

The torque $M$ induced on the two parallel birefringent plates is given by\cite{barash}:

\begin{equation}
M=-{\partial \Omega \over \partial \theta} \cdot S
\label{eqtorque}
\end{equation} 

\noindent where $S$ is the area of the interacting surfaces. The retarded van der Waals (or Casimir-Lifshitz) force is given by:

\begin{equation}
F=-{\partial \Omega \over \partial d} \cdot S
\label{eqforce}
\end{equation} 

Following reference \cite{barash}, it is also possible to show that in the non-retarded limit the torque between two plates made out of slightly birefringent materials (i.e. with  $|{\epsilon_{\|}-\epsilon_{\bot} \over \epsilon_{\bot}}| << 1$) reduces to:

\begin{equation}
M=-{\hbar \overline{\omega} S \over 64 \pi^2 d^2}\cdot \sin(2\theta)
\label{torquesimple}
\end{equation}

\noindent where $\overline{\omega}$ is given by:

\begin{equation}
\overline{\omega}=\int_0^\infty d\xi \int_0^\infty dx \cdot {x (\epsilon_{2\|}-\epsilon_{2\bot})(\epsilon_{1\|}-\epsilon_{1\bot})\epsilon_3^2 e^{-x} \over [(\epsilon_{1\bot}+\epsilon_3)(\epsilon_{2\bot}+\epsilon_3)
-(\epsilon_{1\bot}-\epsilon_3)(\epsilon_{2\bot}-\epsilon_3)e^{-x}]^2}
\label{omegabar}
\end{equation}

\noindent We recall that in equation \ref{omegabar} $\epsilon_{1\|}$, $\epsilon_{1\bot}$, $\epsilon_{2\|}$, and $\epsilon_{2\bot}$ must be evaluated at imaginary frequency $i\xi$.

The integral over $x$ in equation \ref{omegabar} can be solved analytically; the equation for $\overline{\omega}$ reads:

\begin{equation}
\overline{\omega}=-\int_0^\infty d\xi \cdot {(\epsilon_{2\|}-\epsilon_{2\bot})
(\epsilon_{1\|}-\epsilon_{1\bot}) \epsilon_3^2\over (\epsilon_{1\bot}^2-\epsilon_3^2)(\epsilon_{2\bot}^2-\epsilon_3^2)}\cdot \ln \biggl( 1-{(\epsilon_{1\bot}-\epsilon_3)(\epsilon_{2\bot}-\epsilon_3) \over (\epsilon_{1\bot}+\epsilon_3)(\epsilon_{2\bot}+\epsilon_3)}
\biggr) 
\label{simpleomegabar}
\end{equation}

\noindent In the above limits the torque is thus inversely proportional to the second power of $d$ and is proportional to $\sin(2\theta)$. 

In the retarded theory, it is generally not possible to reduce the expression for the torque to a simplified analytical equation. In order to determine the dependence of $M$ on $d$ and $\theta$, it is thus necessary to solve equations \ref{energy} through \ref{eqtorque} numerically. In the next section we will analyze a particular configuration and calculate the torque as a function of $\theta$ at a distance where retardation effects cannot be neglected. 
 
\section{TORQUE IN VACUUM}

We now focus on a particular experimental configuration. We consider a 20 $\mu$m thick, 40 $\mu$m diameter disk made out of either quartz or calcite, kept in vacuum parallel to a large BaTiO$_3$ plate at a distance $d$. It is easy to recognize that to perform the numerical computation of the torque experienced by the disk as a function of $\theta$ and $d$ we only need to know the dielectric functions of the two plates at imaginary angular frequencies $i \xi_n$. The dielectric properties of many materials are well described by a multiple oscillator model (the so-called Ninham-Parsegian representation)\cite{ninham}, which can be written as follows:

\begin{equation}
\epsilon(i \xi )=1+\sum_{j=1}^N{C_j\over 1+\left({\xi\over \omega_j}\right)^2+{g_j \xi \over \omega_j}}
\label{NP} 
\end{equation}

\noindent In equation \ref{NP}, $C_j$ is given by $C_j={2\over \pi}{f_j\over \omega_j}$, where $f_j$ is the oscillator strength and $\omega_j$ is the relaxation frequency multiplied by $2\pi$, while $g_j$ is the damping coefficient of the oscillator. 

For most inorganic materials, only two undamped oscillators are commonly used to describe the whole dielectric function\cite{bergstrom,white}:

\begin{equation}
 \epsilon(i\xi)=1+{C_{IR}\over 1+\left({\xi\over \omega_{IR}}\right)^2}+{C_{UV}\over 1+\left({\xi\over \omega_{UV}}\right)^2}
\label{twoosc}
\end{equation}

\noindent where $\omega_{IR}$ and $\omega_{UV}$ are the characteristic absorption angular frequencies in the infrared and ultraviolet range, respectively, and $C_{IR}$ and $C_{UV}$ are the corresponding absorption strengths. 

The two oscillator model does not always provide a complete description of the dielectric properties of materials; however, in spite of its simplicity, when applied to dispersion effects it usually leads to rather precise results\cite{bergstrom,white}. We have thus used this model for our calculations. The limits of this choice will be discussed in section VII.

The parameters that determine the dielectric properties of quartz, calcite, and BaTiO$_3$ in the limit of the two oscillator model (equation \ref{twoosc}) are listed in table \ref{tabdf}\cite{bergstrom,white}. Figure \ref{figepsi} shows the calculated $\epsilon(i\xi)$.

Using these functions, we have calculated the torque expected for different angles at $d=100$ nm, both in the quartz-BaTiO$_3$ and in the calcite-BaTiO$_3$ configurations. The results obtained for $T=300$ K are reported in figure \ref{qorcvb}. Calculations for smaller temperatures give rise to nearly identical values: at this distance the torque is solely generated by the fluctuations of the electromagnetic field associated with the zero-point energy, because contributions arising from thermal radiation can be neglected\cite{noteontemp}.

The computational data were interpolated using a sinusoidal function with periodicity equal to $\pi$ ($M=a\sin(2\theta)$); its amplitude ($a$) was adjusted by means of an unweighted fit. This curve, reported in figure \ref{qorcvb}, well interpolates the numerical results. 

The maximum magnitude of the torque occurs at $\theta={\pi\over 4}$ and $\theta={3\over 4}\pi$. Comparison of the results obtained with quartz and calcite shows that, as expected, materials with less pronounced birefringent properties (such as quartz with respect to calcite) give rise to a smaller torque. 

Interestingly, the sign of the torque obtained for the quartz-BaTiO$_3$ configuration is opposite to the one obtained for calcite-BaTiO$_3$. The reason for this behavior can be understood from the dependence of the dielectric functions on imaginary frequency (figure \ref{figepsi}). For quartz, $\epsilon_{\|}>\epsilon_{\bot}$ at all frequencies, while for BaTiO$_3$, $\epsilon_{\bot}$ is always larger than $\epsilon_{\|}$. For calcite, the two curves cross at $\xi\simeq 2\cdot 10^{14}$ rad/s; however, the contribution to the torque arising from frequencies below this value are relatively small. Therefore, the largest contribution to the torque comes from angular frequencies in the region where $\epsilon_{\bot}>\epsilon_{\|}$. The minimum zero-point energy corresponds to the situation in which the axes of the dielectric tensors with larger values of $\epsilon$ are aligned. For the quartz-BaTiO$_3$ combination, this situation is reached for $\theta={\pi\over 2}$: the torque is positive from $\theta=0$ to $\theta={\pi\over 2}$ and negative from $\theta={\pi\over 2}$ to $\theta=\pi$. For calcite-BaTiO$_3$, on the contrary, the minimum energy corresponds to $\theta=0$: the sign of the torque is thus reversed with respect to the previous case.

We have also calculated the magnitude of the torque at $\theta={\pi\over 4}$ as a function of the distance between the disk and the plate. From the results reported in figure \ref{mdv} one can clearly verify that it is not possible to infer a single power law dependence that describes the torque at all distances regardless of the choice of the interacting materials. 

For $d=100$ nm, the maximum magnitude of the torque is approximately equal to $5.3\cdot 10^{-20}$ N$\cdot$m for the quartz disk and $7\cdot 10^{-19}$ N$\cdot$m for the calcite disk. In 1936 R. A. Beth performed an experiment where a torsional balance was used to measure the rotation of a macroscopic quartz disk induced by the transfer of angular momentum of light\cite{beth}. He achieved a sensitivity of $\simeq 10^{-17}$ N$\cdot$m\cite{opticaltrap}. It is thus reasonable to ask if similar set-ups with today's improved technology could be used to observe the rotation between the $40$ $\mu$m diameter disk and the BaTiO$_3$ plate induced by virtual photons associated with the zero-point energy . The main difficulty of this experiment, and the main difference with the measurement cited above, is that it is necessary to keep the disk freely suspended just above the other plate at separations where the two surfaces would tend to come into contact. From figure \ref{mdv}, we can estimate that in order to observe the effect the two surfaces should be kept at least at sub-micron distances. One could argue that the plate used in Beth's experiment was much larger than the disk that we have considered so far. Because the torque is proportional to the area of the interacting surfaces, one could use a plate with a much larger diameter. For example, for a 1 cm diameter quartz disk kept parallel to a BaTiO$_3$ plate, the torque is larger than $10^{-17}$ N$\cdot $m for $d\lesssim 1$ $\mu$m. However, other problems would arise in this case. The surface roughness and the curvature of the two birefringent plates should be much smaller than $1 \mu$m over an area of several cm$^2$. This area should also be completely free from dust particles with diameter larger than a few hundreds of nm. Furthermore, one should still design a mechanical set-up to keep the two slabs parallel without compromising the sensitivity of the instrument. We conclude that the use of a torsional balance for the measurement of the torque induced by quantum fluctuations would present several major technical problems. Instead of discussing this possibility, we propose a simpler solution. Let's consider again the microscopic disk described at the beginning of this section. We will show below that the retarded van der Waals force between quartz or calcite and BaTiO$_3$ in liquid ethanol is repulsive; thus, if the disk is placed on top of the plate, this repulsion can be used to counterbalance the weight of the quartz disk. In liquid ethanol, therefore, the disk would float parallel on top of the BaTiO$_3$ plate at a small distance. The static friction between the two birefringent plates would be virtually zero, and the disk would be free to rotate suspended in bulk liquid. If the torque induced by quantum fluctuations does not sensibly decrease after the introduction of liquid ethanol in the gap, and if the equilibrium distance is smaller than a few hundreds nanometer, the configuration proposed should allow the demonstration of the rotation of the disk in a reasonably straightforward experiment.

\section{RETARDED VAN DER WAALS FORCE IN LIQUID ETHANOL}

The force between the disk and the plate was calculated according to equation \ref{eqforce}. The parameters used to determine the dielectric properties of ethanol are reported in table \ref{tabdf}\cite{repulsive1}. The results of the calculations for $\theta={\pi \over 4}$ are reported in figure \ref{force}. Similar calculations were carried out for $\theta=0$ and $\theta={\pi \over 2}$: the results differ from the curves represented in figure \ref{force} by less than 10$\%$.

For distances shorter than a few nanometers the force is attractive. However, at larger distances, where retardation effects start to play an important role in the interaction, the force switches to repulsive. The reason for this behavior can be understood by comparing the dielectric functions of quartz, calcite, BaTiO$_3$, and ethanol, which are reported in the inset of figure \ref{force}. For sake of simplicity, for each birefringent material, we show the average of $\epsilon_{\bot}$ and $\epsilon_{\|}$. The arguments discussed below refer to the average of $\epsilon_{\bot}$ and $\epsilon_{\|}$, but they can be similarly applied to the two principal components of the dielectric tensor separately. For $\xi\lesssim 5\cdot 10^{15}$, we have:

\begin{equation}
\epsilon_{\textrm{quartz}}<\epsilon_{\textrm{ethanol}}<\epsilon_{\textrm{BaTiO$_3$}} \qquad ; \qquad \epsilon_{\textrm{calcite}}<\epsilon_{\textrm{ethanol}}<\epsilon_{\textrm{BaTiO$_3$}} \label{epsicomb}
\end{equation}

\noindent while at higher frequencies $\epsilon_{\textrm{ethanol}}$ is smaller than $\epsilon_{\textrm{quartz}}$, $\epsilon_{\textrm{calcite}}$, and $\epsilon_{\textrm{BaTiO}_3}$. From Lifshitz's theory for isotropic materials, it is possible to show that the force between two plates with dielectric functions $\epsilon_1$ and $\epsilon_2$ immersed in a medium with dielectric function $\epsilon_3$ is repulsive if, for imaginary frequencies, $\epsilon_1<\epsilon_3<\epsilon_2$ or $\epsilon_2<\epsilon_3<\epsilon_1$, and it is attractive in all other cases\cite{ninham,isra}. In our system there is a crossover from $\epsilon_1<\epsilon_3<\epsilon_2$ to $\epsilon_3<\epsilon_1<\epsilon_2$ at $\xi \simeq 5\cdot 10^{15}$ rad/s. The force is thus repulsive at large distances, where low imaginary frequencies give rise to the most important contribution to the force, and attractive for smaller separations, where higher frequencies are more relevant\cite{milonni,ninham,isra}.

The zero-point energy due to electromagnetic quantum fluctuations depends on the distance between the two interacting plates. If the condition $\epsilon_1<\epsilon_3<\epsilon_2$ (or $\epsilon_2<\epsilon_3<\epsilon_1$) is satisfied, the zero-point energy per unit area is smaller at larger separation, which means that it is energetically more favorable for the liquid to stay inside the gap rather than outside. As a consequence, the net force between the plates is repulsive. For detailed energy balance considerations and a more rigorous proof of this statement, we refer the reader to the so-called \textit{Hamaker theorem}, discussed in \cite{ninham}.

The net weight of the disk immersed in ethanol is given by:
\begin{equation}
F_{gr}=-S\cdot h \cdot (\rho_{\textrm{disk}}-\rho_{\textrm{ethanol}}) \cdot g
\label{gravit}
\end{equation}

\noindent where $S$ is the surface of the disk, $h$ is its thickness, $\rho_{\textrm{disk}}$ is the mass density of either quartz (2643 kg/m$^3$) or calcite (2760 kg/m$^3$), $\rho_{\textrm{ethanol}}$ is the mass density of ethanol (789 kg/m$^3$), and $g=9.81$ m/s$^2$. The arrow on the curves reported in figure \ref{force} indicates the distance at which this force is counterbalanced by the retarded van der Waals repulsion. In both the cases under investigation, the equilibrium separation is about 100 nm. This distance can be tailored to the experimental needs by changing the thickness of the disk.

\section{TORQUE BETWEEN PLATES IN LIQUID}

We have calculated the expected torque between the disk and the plate in liquid ethanol as a function of angle and distance. We have verified that also in this case the torque varies as $\sin(2\theta)$ and has maximum magnitude at $\theta={\pi \over 4}$ and $\theta={3\pi \over 4}$. The sign of the rotation that one should observe using a quartz disk is opposite with respect to what is expected for a calcite disk, and temperature corrections are negligible. 

Figure \ref{mde} shows the calculated magnitude of the torque at $\theta={\pi\over 4}$ for different values of $d$. At $d=100$ nm the torque is smaller by a factor of 2 with respect to the case of vacuum. 

Note that at short distances the torque in ethanol is actually larger than in vacuum (figure \ref{mde}). At present, we do not have an intuitive explanation of this phenomenon. 

\section{PROPOSED EXPERIMENT}

A schematic view of the proposed experimental set-up is shown in figure \ref{setup}. A 40 $\mu$m diamater, 20 $\mu$m thick disk made out of quartz or calcite is placed on top of a BaTiO$_3$ plate immersed in ethanol.

The optical axes of the birefringent crystals are oriented as in figure \ref{Barash_config}. According to the arguments in the previous section, the disk should levitate approximately 100 nm above the plate and should be free to rotate in a sort of \textit{frictionless bearing}. A 100 mW laser beam can be collimated onto the disk to rotate it by the transfer of angular momentum of light. A shutter can then block the beam to stop the light-induced rotation. The position of the disk can be monitored by means of a microscope objective coupled to a CCD-camera for imaging.  

Using the laser, one can rotate the disk until $\theta={\pi \over 4}$. Once the laser beam is shuttered, the disk is free to rotate back towards the configuration of minimum energy according to the following equation (see Appendix):

\begin{equation}
I\cdot \ddot{\theta}+{\pi \over 2}\cdot {R^4\over d} \cdot \eta \cdot \dot{\theta}=a\cdot \sin(2\theta)
\label{thetat}
\end{equation}

\noindent where $R$ is the radius of the disk, $I$ is its momentum of inertia, $\eta$ is the viscosity of ethanol ($1.2\cdot 10^{-3}$ Ns/m$^2$), $d$ is the distance between the disk and the plate, and $a\sin(2\theta)$ is the torque due to quantum fluctuations. For an estimate of the time evolution, we have determined $a$ from figure \ref{mde} and solved equation \ref{thetat} for $\theta(t=0)={\pi \over 4}$ and $\dot{\theta}(t=0)=0$. The results are reported in the inset of figure \ref{setup}. Note that the rotation is overdamped in both cases: the disk moves asymptotically and monotonically towards the equilibrium position. For the calcite disk, easily measurable rotations should be observed within a few minutes after the laser beam shutter is closed. The quartz disk would rotate much slower, and it is questionable whether its rotation could be detected or not. However, it is worth to stress that the set-up presented above is not yet optimized. Disks with different dimensions and geometries might rotate faster. Suitably engineered samples could also result in more favorable experimental configurations. For example, a thick layer of lead could be deposited on a portion of the disk to make it heavier: the disk would then float at a smaller distance, where the magnitude of the driving torque would be larger. Furthermore, the use of a different liquid with optical properties similar to ethanol but with a smaller viscosity would significantly increase the angular velocity of the disk. Finally, a more sophisticated optical set-up could be implemented for the measurement of small rotations.

One could argue that it might be difficult to distinguish the cause of the rotation of the disk from other effects that could mimic the phenomenon under investigation. Surface roughness, charge accumulation, and liquid motion could be typical reasons. However, there are two defining properties that should help  experimenters rule out spurious effects: (i) the torque induced by quantum fluctuations has periodicity $\pi$, and (ii) the sign of the torque depends on whether the experiment is performed with a quartz or a calcite disk.  For example, a calcite disk would rotate clockwise if initially positioned at ${\pi\over 4}$ or ${5\over 4}\pi$, and anti-clockwise if initially positioned at ${3\over 4}\pi$ or ${7\over 4}\pi$. For a quartz disk, one would obtain the opposite behavior. An experimental observation of the dependence of the rotation direction on the initial position and on the choice of the interacting materials would thus indicate that the disk is solely driven by the quantum fluctuations of the electromagnetic field. As an additional proof, the experiment could be repeated by placing the disk over a non-birefringent plate with $\epsilon(i\xi) > \epsilon_{\textrm{ethanol}}(i\xi)$. The retarded van der Waals force would still be repulsive, but there would be no torque induced by quantum fluctuations. 

\section{PRECISION AND LIMITS OF PREVIOUS CALCULATIONS}

Calculations were performed using Mathematica (Version 5, Wolfram Research). Although integrals and summations are estimated to be exact with a level of accuracy $\lesssim 1\%$, the overall results cannot be considered equally precise. The model used for the dielectric function of the materials (equation \ref{twoosc}) is in fact relatively unprecise. To give a sense of how much the model might affect the results, we will reconsider below the configuration of quartz-ethanol-BaTiO$_3$. In our previous calculations we assumed $\omega_{IR}=0.85\times 10^{14}$. This value is the average of the two values available in the literature\cite{bergstrom}, $\omega_{IR,1}=0.7\times 10^{14}$ and $\omega_{IR,2}=1\times 10^{14}$. At $d=100$ nm and $\theta={\pi\over 4}$, both the torque and the Casimir-Lifshitz force that one obtains using either $\omega_{IR,1}$ or $\omega_{IR,2}$ differs from the results obtained with the average by less than 14$\%$. This discrepancy is due to the fact that the dielectric function of BaTiO$_3$ is very large in the infrared. As a consequence, a large contribution to the summation in equation \ref{energy} comes from the first few terms (i.e. for small $n$), which correspond to relatively large wavelengths. This means that if the model is not accurate enough in that region, larger errors can be introduced in the calculation. 

Although it is obvious that in order to compare experiment to theory a deeper knowledge of the dielectric properties of the materials is needed, the use of slightly different values of the parameters in $\epsilon(i\xi)$ does not significantly influence the order of magnitude of our results. 

\section{CONCLUSIONS}

We have performed detailed numerical calculations of the mechanical torque between a 40 $\mu$m diameter birefringent disk, made of quartz or calcite, and a BaTiO$_3$ birefringent plate. At separations of the order of a few hundreds of nanometers, the magnitude of the torque is of the order of $10^{-19}$ N$\cdot$m. We have shown that a demonstration of the effect could be readily obtained if the birefringent slabs were immersed in liquid ethanol. In this case the disk would float on top of the plate at a distance where the repulsive retarded van der Waals force balances gravity, giving rise to a mechanical bearing with ultra-low static friction. The disk, initially set in motion via transfer of angular momentum of light from a laser beam, would return to its equilibrium position solely driven by the torque arising from quantum fluctuations. 

\section{ACKNOWLEDGEMENTS}

Enlightening discussions with L. Levitov and H. 
Stone are gratefully acknowledged.

This work was partially supported by NSEC (Nanoscale Science and Engineering Center), under NSF contract number PHY-0117795.

\section{APPENDIX}

Consider a disk of radius $R$ rotating at angular velocity $\dot{\theta}$ parallel to a surface at distance $d$ and immersed in a liquid with viscosity $\eta$. Inside the gap, the velocity $u$ of the liquid is described by the following equation\cite{tritton}:

\begin{equation}
u(r,\Phi,z)=\dot{\theta} \cdot r \cdot {z\over d}
\label{app1}
\end{equation}

\noindent where we have introduced cylindrical coordinates $\{ r,\Phi,z\}$ in the reference system reported in figure \ref{Barash_config}. The stress induced by the viscosity of the liquid on an infinitesimal area of the disk at coordinates $\{ r,\Phi,d\}$ is given by:

\begin{equation}
\tau(r,\Phi)=\eta\cdot {\partial u(r,\Phi,z) \over \partial z}|_{(z=d)}
\label{app2}
\end{equation}

\noindent The total torque due to the viscosity of the liquid then reads: 

\begin{equation}
D=\int_0^{2\pi}\textrm{d}\Phi\int_0^{R}\tau(r,\Phi)\cdot r \cdot r\textrm{d}r={\pi \over 2} \cdot {R^4\over d} \cdot \eta \cdot \dot{\theta}
\label{app3}
\end{equation}

\noindent which leads to equation \ref{thetat}.

It is interesting to note that the drag is dominated by the torque due to the liquid inside the gap. To estimate the contribution to the drag of the surrounding liquid, one can calculate the drag expected if the disk were rotating in bulk, far away from any other surfaces. In this case, the torque is given by $D={32\over 3}\eta R^3 \dot{\theta}$ (see article cited in note \cite{opticaltrap}), which, for the configuration chosen in our experiment, is a factor $\simeq 30$ smaller than the one expected from equation \ref{app3}.
  
Note also that thermal fluctuations do not contribute to the rotation as long as the disk is far away from the equilibrium position. The potential energy due to quantum fluctuations is in fact much larger than $k_BT$ for angles larger than a few degrees.

\pagebreak
\newpage

\begin{table}
\begin{tabular}{|cc|c|c|c|c|c|}
\hline
& & $C_{IR}$ & $C_{UV}$ & $\omega_{IR}$ (rad/s) & $\omega_{UV}$ (rad/s) \\ \hline
Quartz & $\|$ & 1.920 & 1.350 & $2.093 \cdot 10^{14}$ & $2.040 \cdot 10^{16}$ \\ 
 & $\bot$ & 1.960 & 1.377 & $2.093 \cdot 10^{14}$ & $2.024 \cdot 10^{16}$ \\ \hline
Calcite & $\|$ & 5.300 & 1.683 & $2.691 \cdot 10^{14}$ & $1.660 \cdot 10^{16}$ \\ 
& $\bot$ & 6.300 & 1.182 & $2.691 \cdot 10^{14}$ & $2.134 \cdot 10^{16}$ \\ \hline
BaTiO$_3$ & $\|$ & 3595 & 4.128 & $0.850 \cdot 10^{14}$ & $0.841 \cdot 10^{16}$ \\ 
& $\bot$ & 145.0 & 4.064 & $0.850 \cdot 10^{14}$ & $0.896 \cdot 10^{16}$ \\ \hline
Ethanol &  & 23.84 & 0.852 & $6.600 \cdot 10^{14}$ & $1.140 \cdot 10^{16}$ \\ \hline
\end{tabular}
\caption{Values of the parameters used to determine the dielectric function of the materials.\label{tabdf}}
\end{table}

.
\pagebreak
\newpage
\begin{figure}[t]
\caption{Sketch of the system under investigation. Two birefringent parallel plates with in plane optical axis are placed in close proximity. The $z$-axis is chosen to be orthogonal to the plates. $\theta$ is the angle between the two optical axes.\label{Barash_config}}
\end{figure}

\begin{figure}[t]
\caption{Dielectric functions of the materials used in our calculations as a function of imaginary angular frequency. Solid lines and dotted lines represent, respectively, $\epsilon_{\|}$ and $\epsilon_{\bot}$ of the birefringent slabs: (a) quartz, (b) calcite, (c) Barium Titanate. \label{figepsi}}
\end{figure}

\begin{figure}[t]
\caption{Calculated torque as a function of $\theta$ between a 40 $\mu$m diameter disk made of quartz (a) or calcite (b) and a Barium Titanate plate separated in vacuum by a distance $d=100$ nm. The lines represent a fit of the numerical results with $M=a\sin(2\theta)$. \label{qorcvb}}
\end{figure}

\begin{figure}[t]
\caption{Calculated torque as a function of plate separation between a 40 $\mu$m diameter disk made of quartz ($\bullet$) or calcite ($\times$) and a Barium Titanate plate in vacuum at $\theta={\pi \over 4}$.  \label{mdv}}
\end{figure}

\begin{figure}[t]
\caption{Calculated retarded van der Waals force as a function of plate separation for a 40 $\mu$m diameter disk made of quartz (a) or calcite (b) and a plate of Barium Titanate immersed in ethanol, calculated for $\theta={\pi\over 4}$. The arrow represents the distance at which the retarded van der Waals repulsion is in equilibrium with the weight of the disk. Insets: Dielectric functions of the interacting materials. For the three birefringent samples, the curve represents the average of $\epsilon_{\bot}$ and $\epsilon_{\|}$. \label{force}}
\end{figure}

\begin{figure}[t]
\caption{Calculated torque as a function of plate separation between a 40 $\mu$m diameter disk made of quartz (a) or calcite (b) and a Barium Titanate plate immersed in ethanol ($\bullet$) or vacuum ($\circ$) for $\theta={\pi \over 4}$.
\label{mde}}
\end{figure}

\begin{figure}[t]
\caption{A sketch of the experimental set-up proposed in this paper. Inset: calculated value of the angle between the optical axes of the two birefringent crystals as a function of time.
\label{setup}}
\end{figure}

\pagebreak

\begin{figure}[t]
\epsfxsize=\textwidth
\epsfbox{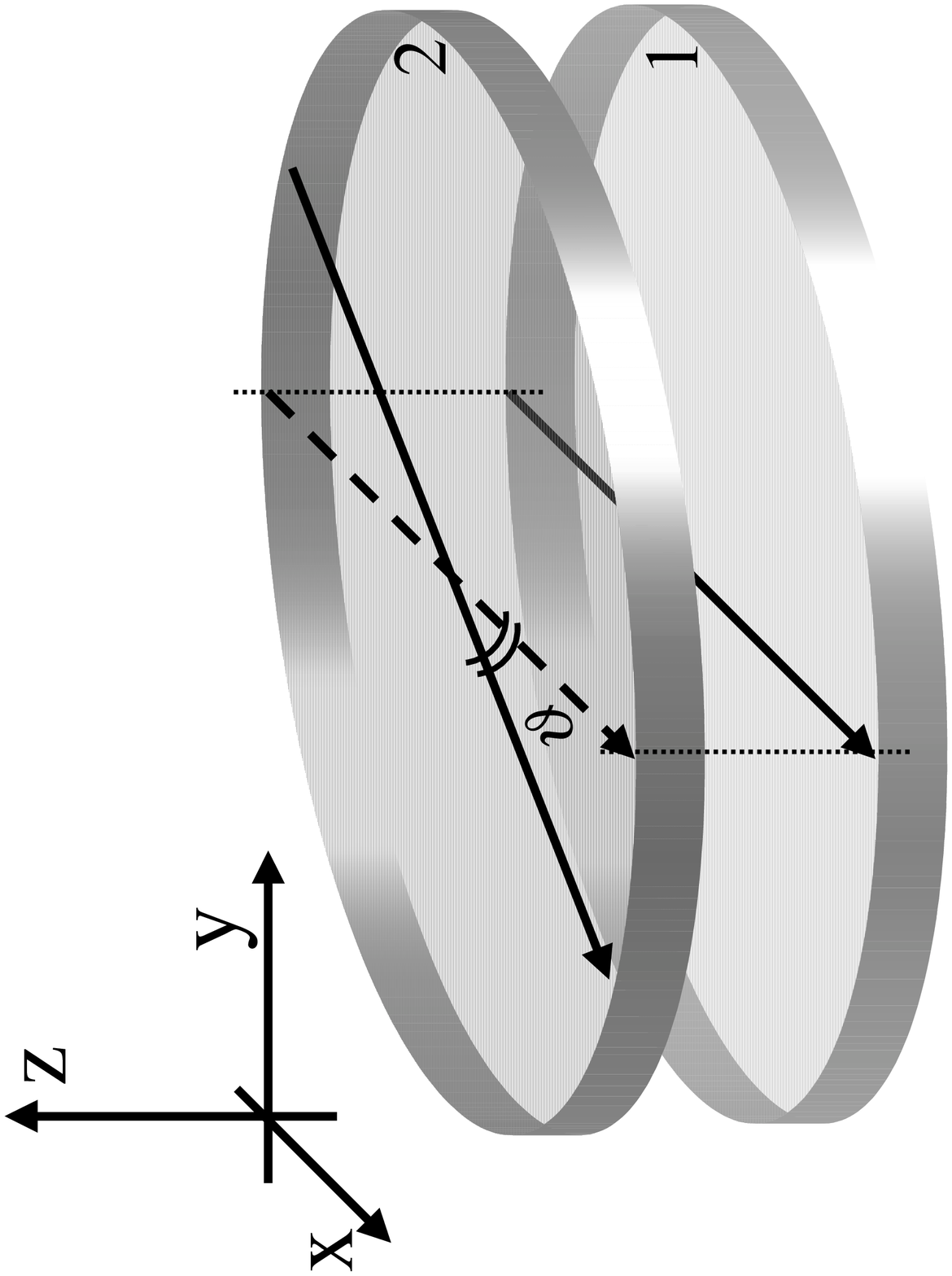}
\end{figure}
\newpage

\begin{figure}[t]
\epsfxsize=\textwidth
\epsfbox{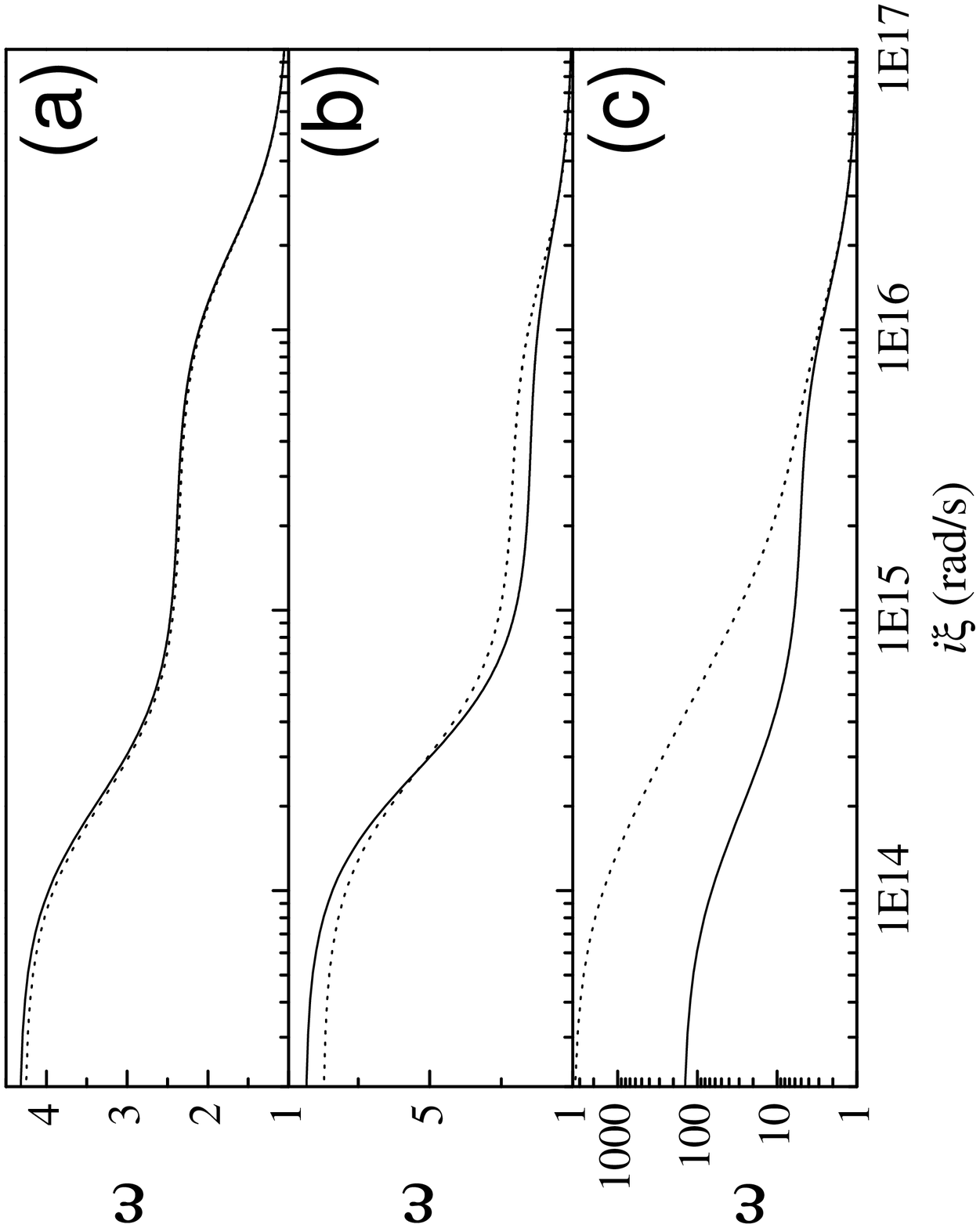}
\end{figure}
\newpage

\begin{figure}[t]
\epsfxsize=\textwidth
\epsfbox{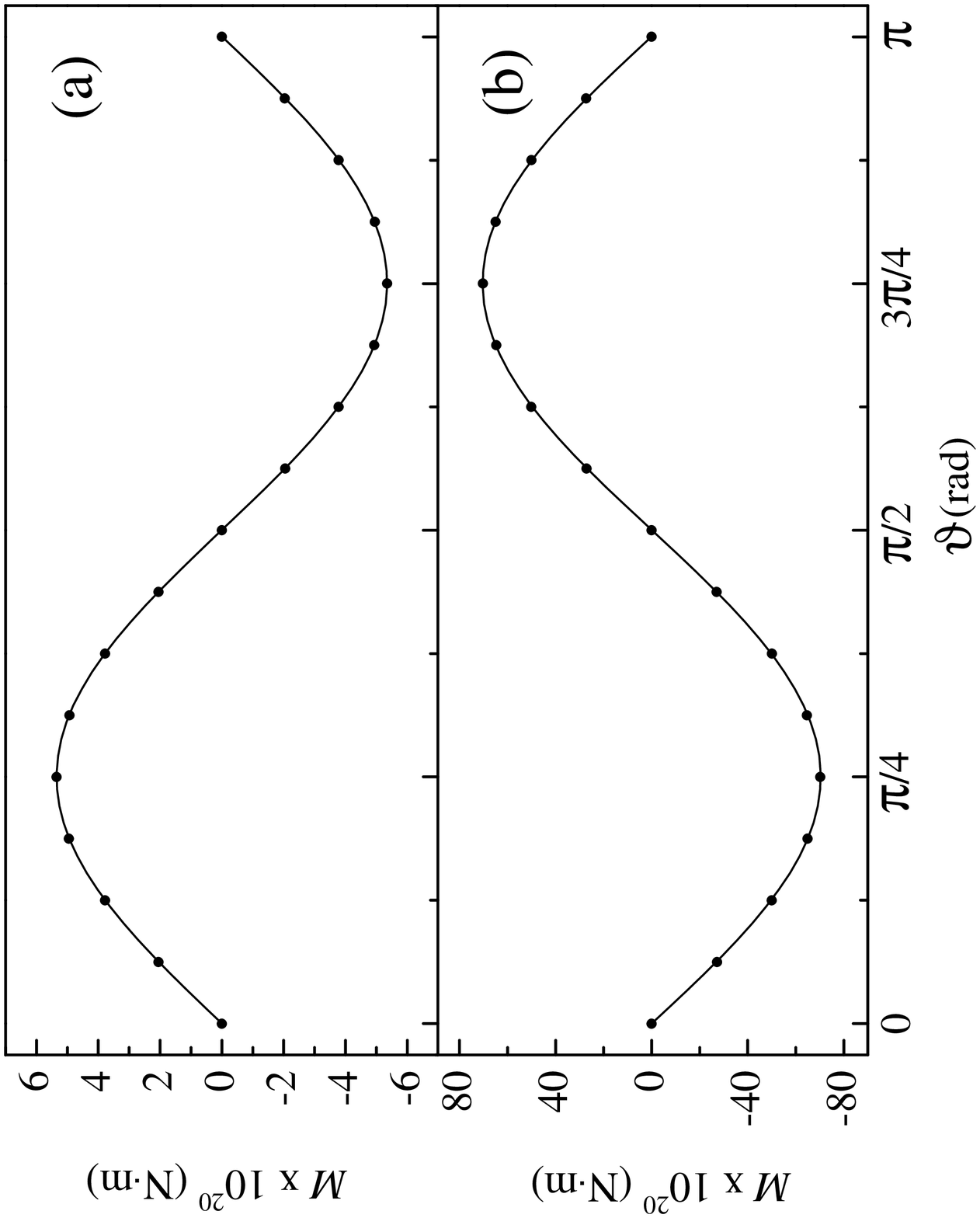}
\end{figure}
\newpage

\begin{figure}[t]
\epsfxsize=\textwidth
\epsfbox{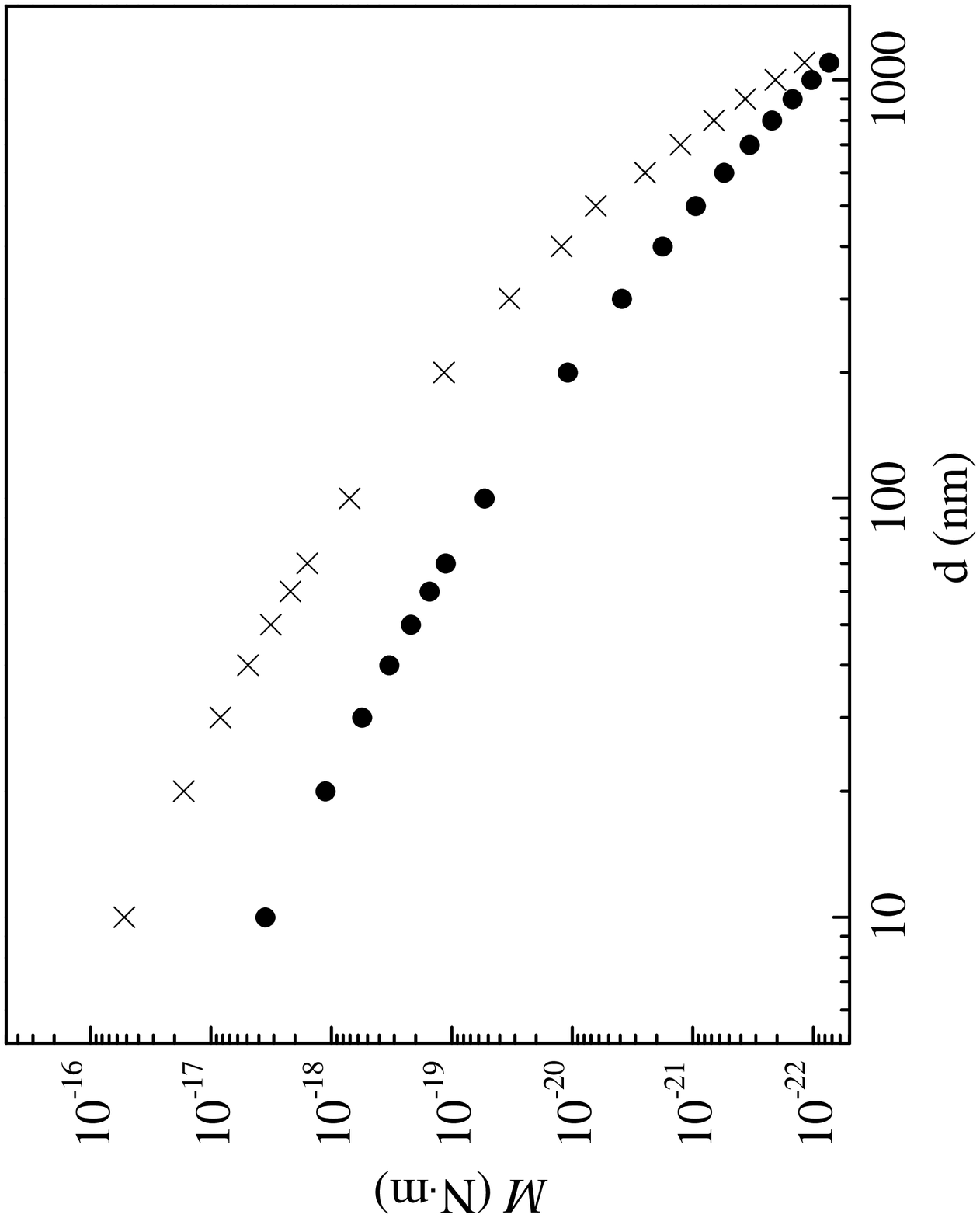}
\end{figure}
\newpage

\begin{figure}[t]
\epsfxsize=\textwidth
\epsfbox{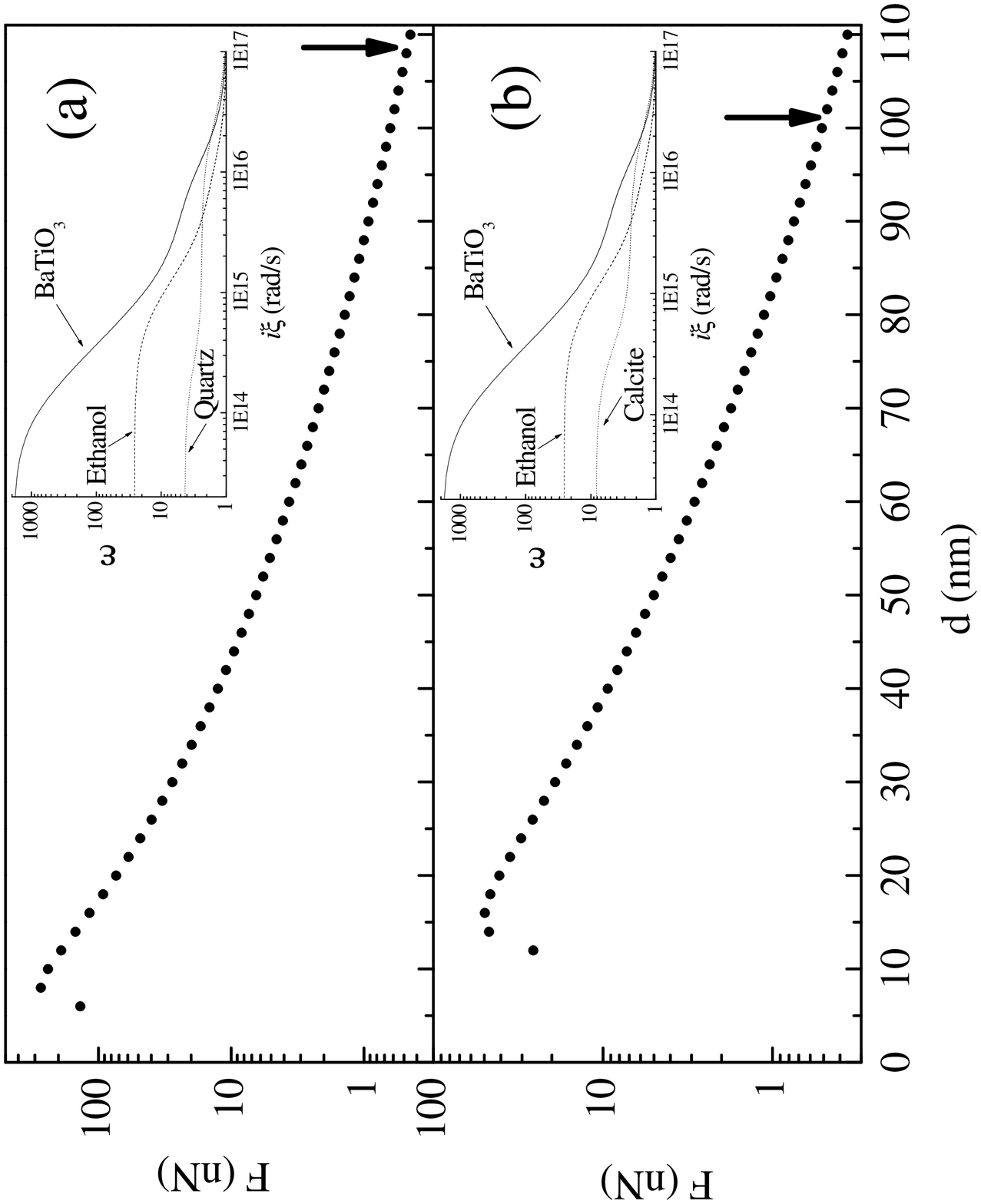}
\end{figure}
\newpage

\begin{figure}[t]
\epsfxsize=\textwidth
\epsfbox{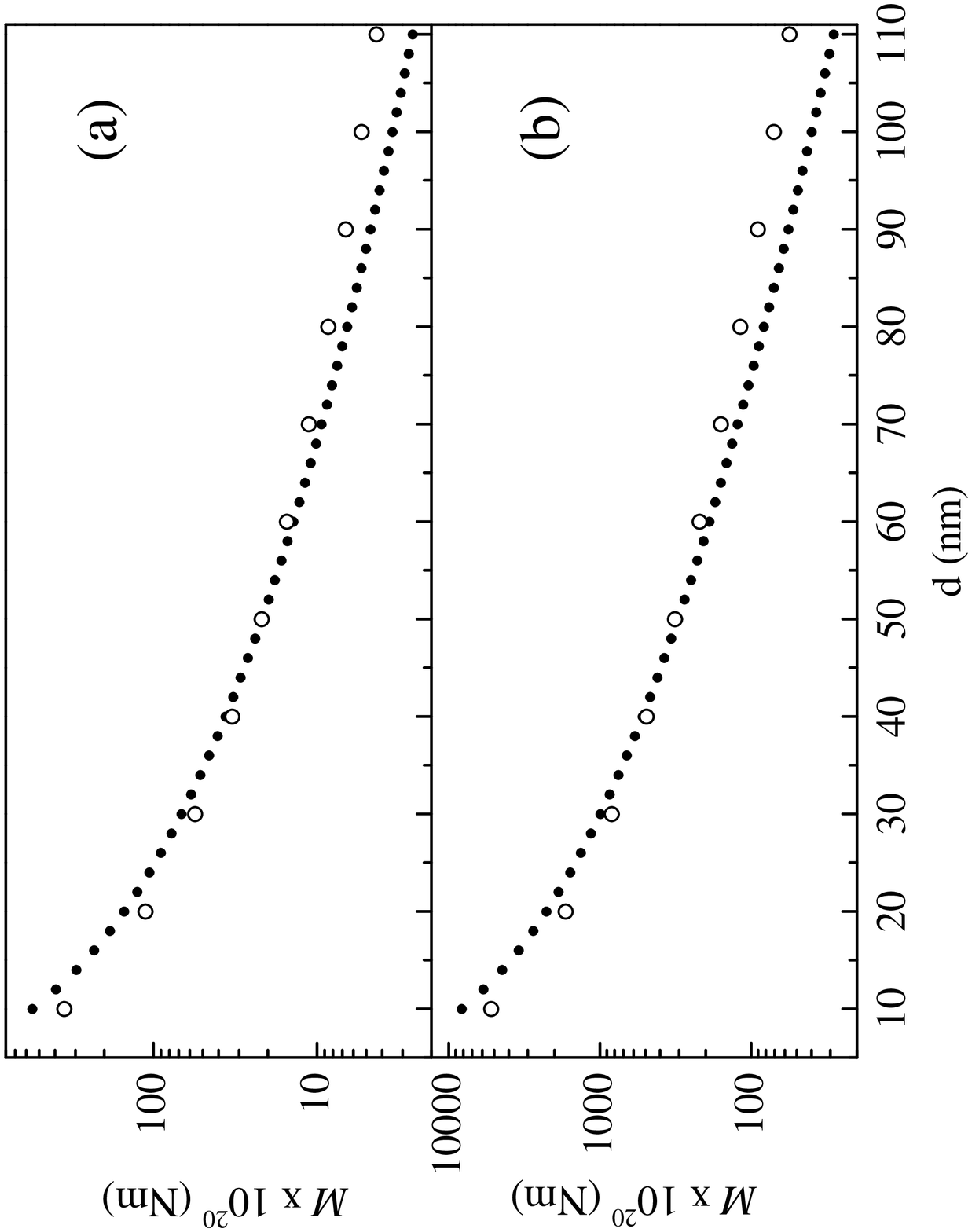}
\end{figure}
\newpage

\begin{figure}[t]
\epsfxsize=\textwidth
\epsfbox{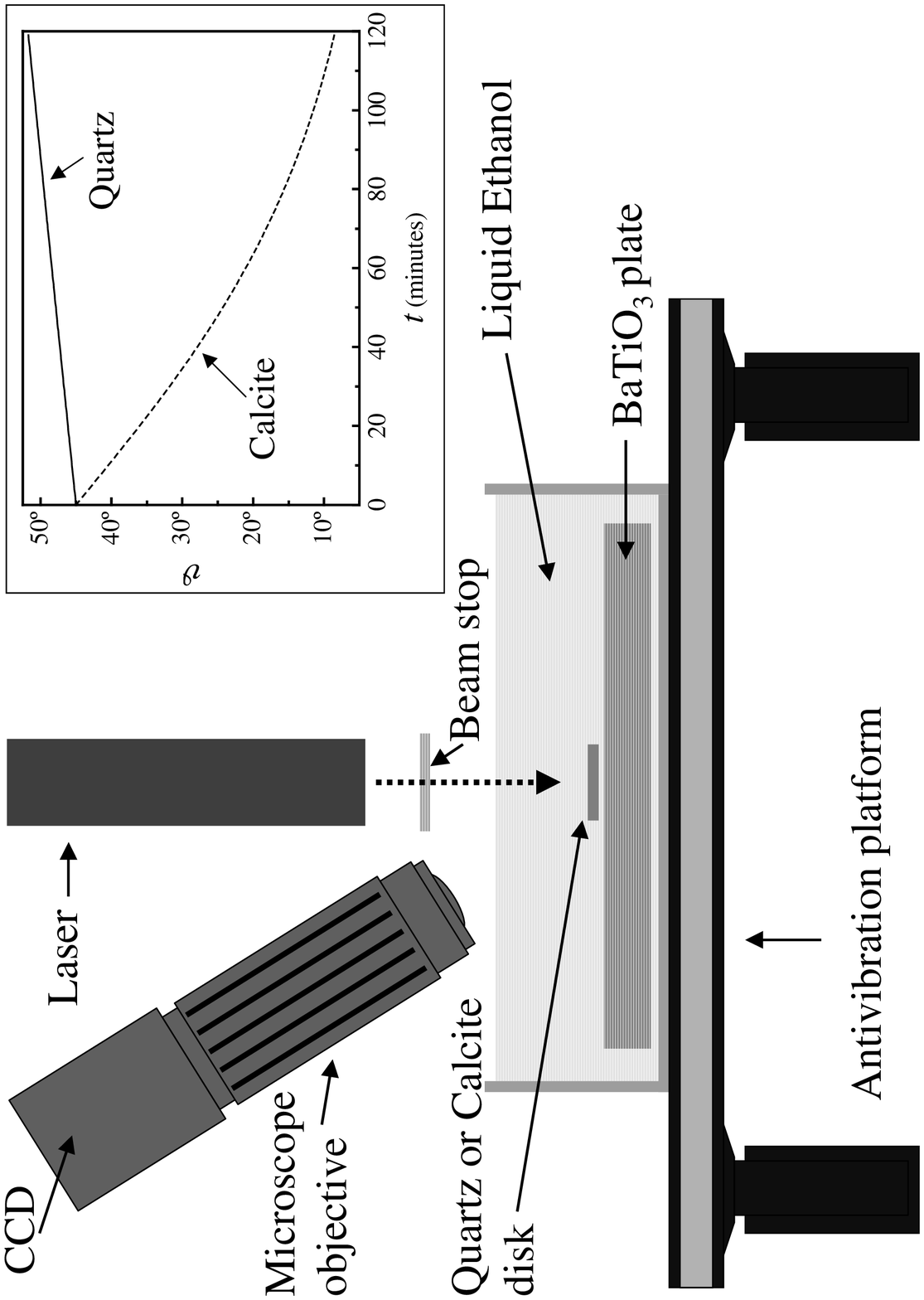}
\end{figure}
\newpage


\begin{references}
\bibitem{noteonlif2} In Lifshitz's theory, the magnetic properties of materials are neglected, and the magnetic permeability is assumed to be equal to 1 at all frequencies.
\end{references}
\end{document}